PHOTOSYNTHESIS ON A PLANET ORBITING AN M DWARF: ENHANCED EFFECTIVENESS DURING FLARES

Short title: Flare-enhanced photosynthesis

Authors:    D. J. Mullan[1] and H. P. Bais[2]

Addresses:  [1]Bartol Research Institute, Dept of Physics and Astronomy, University of Delaware, Newark DE 19716

[1]Dept of Plant and Soil Science, University of Delaware, Newark DE 19716

E-mail address: mullan@udel.edu

ABSTRACT

On planets near M dwarfs, oxygenic photosynthesis (PS) will occur with an effectiveness which depends on the supply of visible photons with wavelengths between 400 and 700 nm. In this paper, we quantify the effectiveness of PS in two contexts which are relevant for M dwarfs. First, using photons from an M dwarf in its *quiescent* (non-flaring) state, we find that PS on an M dwarf planet in the HZ of its parent star is less effective than on Earth by a factor of 10 for a flare star with mid-M spectral type. For a flare star with late-M spectral type, PS effectiveness is smaller than on Earth by a factor of 100 or more. Second, using photons which are incident on the HZ planet *during flares*, we find that PS effectiveness can increase by factors of 5-20 above the quiescent values. In the case of a flare star with mid-M spectral type, we find that the PS effectiveness during a flare can increase up to as much as 50-60% of the values on Earth. However, for a late-M flare star, even during flares, the PS effectiveness remains almost one order of magnitude smaller than on Earth. We suggest that for biological processes on M dwarf planets, the stellar activity cycle may replace the orbital period as the "year".

1. **Introduction**

Planets in orbit in the habitable zone (HZ) around M dwarfs are of great interest as regards the search for life. For a recent discussion of some of the extensive literature on this topic, see (e.g.) Shields et al (2016), Cui et al (2017), and Ranjan et al (2017). In this paper, we are interested in one particular aspect of exoplanetary life (photosynthesis: hereafter PS), and how this is affected by variations in the radiative output from the parent star when that star is an M dwarf.

A well-known aspect of M dwarfs is that most of them are subject to flares, i.e. temporary events during which the star brightens by some (variable) amount and then, over the **range** of a finite time, returns to its quiescent level (e.g. Gershberg 2002). **When we refer to "the" brightness of an M dwarf during a flare, we are referring to the intensity of stellar radiation integrated over the entire visible disk of the star. This integrated radiation includes the quiescent radiation from the undisturbed (non-flaring) portions of the surface, plus more intense localized radiation from an individual active region in which the flare has actually occurred. A nearby planet will receive the combined spectrum of flare plus quiescent radiation.** For present purposes, the most salient aspect **of the radiation received by a planet during flares** is that the spectrum of the radiation extends over a broad range of wavelengths. In a recent paper, Ranjan et al (2017) focused on flare spectra in the ultraviolet (UV), and demonstrated that the dose rates of flare UV photons in a certain range of wavelengths (200-300 nm) are useful in enhancing pre-biotic chemistry.

In the present paper, we focus on a portion of the flare spectrum which was not discussed by Ranjan et al., namely, the *visible* range of wavelengths (400-700 nm). We are interested here in how the visible flare photons contribute to oxygenic photosynthesis (PS). While it is well known that PS ultimately provides energy (either directly or indirectly) for most life on Earth, PS is also of interest as regards the search for life on exoplanets. The reason is that PS produces planetary-scale biosignatures, including distinctive reflectance spectra (the "red edge") of the PS organisms, as well as identifiable biogenic gaseous products (including $O_2$, $O_3$, $CO_2$) (e.g. Heath et al 1999; Des Marais et al 2002; Kiang et al 2007; Seager 2014; Berdyugina et al 2016).

A planet orbiting an M dwarf will find itself, during a flare, subject to fluxes of PS-driving photons which are larger than those emitted by the quiescent star. In this paper, we quantify the enhancement that is expected to occur in PS during a flare compared to that which occurs in the quiescent state. To do this, we convolve the wavelength distributions of the two factors which play a role in PS: (i) the photons which are emitted by the star, and (ii) the absorbance of the material which gives rise to PS. In Section 2, we provide details of factor (i). In Section 3, we discuss **the properties of chlorophyll which are pertinent to the present paper: in particular, we present in Section 3.3 in graphical format, the wavelength-dependent details** of factor (ii). Results of the convolution of the two factors are presented in Section 4 for two particular M dwarfs for which suitable spectra exist in the literature. Some further relevant properties of flares are summarized in Section 5. Discussion and Conclusions are presented in Sections 6 and 7.

## 2. The Radiation Environment in the Habitable Zone

Planets in HZ around stars of different spectral type are situated at different radial distances from the parent star. The definition of HZ we use here is that the mean surface temperature of the planet is 288 K, i.e. the radiation which is incident at the top of the atmosphere has (more or less, ignoring differences in albedo and greenhouse trapping) the same energy flux as the Earth receives from the Sun when integrated over all wavelengths. Thus, the integrated energy flux incident on the top of the planetary atmospheres we consider here is of order 1370 W m$^{-2}$.

Planets in the HZ around parent stars of different spectral types receive photon fluxes which are strongest at different wavelengths. E.g., near an F star, more energy is incident at short wavelengths, while near an M dwarf, more energy is incident at long wavelengths. And yet, by definition, the **wavelength-integrated flux** must be the same in all cases. As a result, when spectra of stars of different spectral types are plotted in a single graph such that all lead to planetary temperatures of 288 K, there will be a "cross-over wavelength" for each spectral type (e.g. Kiang et al 2007; Shields et al 2016). **By the term "cross-over wavelength", we mean the wavelength where two spectral distributions have the same flux.** We shall use the measured flux at the cross-over wavelength to normalize the spectra, thereby enabling a meaningful comparison between PS on Earth and PS on planets around M dwarfs.

In this paper, we focus on two particular flare stars for which spectra are available in the literature over appropriate wavelength ranges. One flare occurred on a star with mid-M spectral type (YZ CMi), and the second on a star with late-M spectral type (2MASSW J0149090+295613: hereafter J0149). A first step in our work is therefore to identify the cross-over wavelengths for these two M dwarfs. To do that, we need to start with the radiation emitted by the Sun as a function of wavelength.

### 2.1. *The Radiation Environment Incident on Earth*

Neckel and Labs (1984: hereafter NL) reported measurements of the Sun's mean intensity (in units of W cm$^{-2}$ ster$^{-1}$ Å$^{-1}$) which are incident on the Earth's atmosphere at wavelengths between 330 and 1250 nm. For present purposes, we have extracted (from their Figs. 4 and 5) values of the mean intensity at selected wavelengths throughout that range. Inserting an angular size of the Sun of 6.67 x 10$^{-5}$ steradians, we convert the NL results to irradiances, and plot the results in units of W cm$^{-2}$ μm$^{-1}$ in Fig. 1 (short dashed line). Integrating the flux from 330 to 1250 nm, NL obtained a flux of 1059.7-1062.4 W m$^{-2}$ depending on corrections to the continuum. According to NL, measurements made by other researchers at longer wavelengths add 271.3-276.8 W m$^{-2}$ to the integrated flux, while the inclusion of UV wavelengths adds 37.0-38.4 W m$^{-2}$. Combining these, NL reported a range of values for the "solar constant", namely, 1369-1378 W m$^{-2}$ at the top of Earth's atmosphere.

More recently, due to re-calibrations of space-borne detectors, the IAU has adopted a "nominal" solar constant of 1361 W m$^{-2}$ (Prsa et al 2016). In view of the 1-2% ranges in the numerical values for the "solar constant", and since the detailed NL data are especially useful in calculating PS effectiveness, we adopt a "mean solar constant" of 1370 W m$^{-2}$ (i.e. lying within the NL range) when we calculate the integrated fluxes of our 2 M dwarfs.

It is noteworthy that some 25% of the Sun's integrated flux occurs at wavelengths longer than 1.25 µm microns. For M dwarfs, the percentages of the integrated flux beyond 1.25 µm will be even larger than for the Sun: therefore, in order to obtain reliable integrated fluxes for M dwarfs, we need to have access to flux measurements at wavelengths extending to many microns.

### 2.2. Determining the cross-over wavelength for a mid-M Star

In order to determine the cross-over wavelength for our mid-M star (YZ CMi), it would ideally be necessary to have access to the spectrum of YZ CMi at wavelengths extending from the UV to far out into the infrared. However, the most extensive spectra we have been able to identify for YZ CMi span wavelengths only from 350 nm to 900 nm (Kowalski 2012). The wavelength range from 0.35 to 0.9 µm is too narrow to provide a reliable value for the integrated flux. However, a spectrum of another flare star (AD Leo) extending from 0.115 µm to 173 µm is available on-line (Cohen, 2017). According to the Simbad database, the spectral type of AD Leo is M4Vae, while YZ CMi is listed as M4.0Ve: these spectral types are so similar to each other that we will adopt the extensive AD Leo spectrum as a proxy for YZ CMi.

Integrating over the entire AD Leo spectrum, and applying a distance of 4.87 pc, we find that a planet at a distance of 1 AU from AD Leo would receive an integrated input flux of 32 W m$^{-2}$ : therefore, if a planet in orbit around AD Leo is to receive 1370 W m$^{-2}$ (and therefore lie in the HZ), the planet must be at a distance of √(32/1370) = 0.15 AU from AD Leo. (With a stellar radius of order 0.4R(sun), the HZ planet is at a distance of order 80 stellar radii from AD Leo.) The input irradiance spectrum to such a planet is illustrated by crosses in Fig. 1: the units are the same as those used by NL for the case of sunlight arriving at Earth. Inspection of Fig. 1 shows that the cross-over for AD Leo and the Sun occurs at a wavelength of 0.8 µm.

In order to discuss PS on a planet in the HZ around YZ CMi, we shall assume that the cross-over wavelength is 800 nm.

### 2.3. Determining the cross-over wavelength for a late-M Star

Liebert et al (1999) have reported on J0149 which underwent a "spectacular flare". The visible spectra reported by Liebert et al, in quiescence and during the flare, span the wavelength range from 630 nm to 1000 nm.

In order to obtain an integrated flux and thereby identify the cross-over wavelength for this star, we need to supplement the data from Liebert et al with data at both shorter and longer wavelengths. As regards **extrapolations to** wavelengths that are shorter than 630 nm, **we could adopt a variety of approaches to extrapolating the spectrum: e.g. scale the observed spectrum of an M dwarf, or use a Planck function.**

**As regards scaling the observed spectrum of an M dwarf in quiescence,** we note that the quiescent spectrum of YZ CMi (see Fig. 3 below) is observed to decrease almost monotonically towards shorter wavelengths, falling essentially to zero at 400 nm. **In principle, we might consider adding supplemental spectral information for J0149 at wavelengths less than 630 nm by scaling the observed spectrum of** the quiescent spectrum of YZ CMi: however, this is not reliable, because the spectral type of J0149 (M9.5-9.7) is so much later than the spectral type of YZ CMi (M4.0). As a result, the spectral features in YZ CMi between 400 and 630 nm may have little to do with the spectral features which occur in the J0149 spectrum in the same wavelength range. To illustrate this point, we may cite the results of Rajpurohit et al (2013), who have illustrated the optical-to-red spectral energy distributions for M dwarfs from M0 to M9.5: this range of spectral types extends late enough to overlap with one of the spectral types assigned to J0149. In principle, we would like to examine the wavelength range from 630 nm to 400 nm: unfortunately this is not possible because the wavelength range shown by Rajpurohit et al (in their Figure 2) extends only to 520 nm. But using the data which are available in their Fig. 2, we see that in an M4 star, the (quiescent) spectrum between 630 and 520 nm certainly contains spectral features which are plainly visible. On the other hand, in an M9.5 star, in the same wavelength range, the spectrum contains no visible features: therefore, we consider that scaling the spectrum from an M4 star in the range 630-520 nm would not necessarily yield a reliable measure of the spectrum of an M9.5 star in the range 520-630 nm. *A fortiori* , we believe that it would not be helpful to scale the YZ CMi spectrum over the range 400-630 nm and apply the scaled results to J0149.

As regards using a Planck function in order to extrapolate from 630 to 400 nm, we note that, because of the presence of increasingly strong molecular absorptions at lower temperatures, the spectra of M dwarfs in the optical-to-red region depart more and more from a Planck spectrum as we examine later and later spectral types. This can be seen also in the paper by Rajpurohit et al (2013: their Fig. 3). In view of this, we consider that there would be little to be gained by using a Planck function for the extrapolation of J0149 to wavelengths shorter than 630 nm.

In view of the difficulties associated with an extrapolation from 630 to 400 nm using either a scaling from the observed spectrum of a hotter M dwarf or a Planck function, we have chosen a less rigorous approach: we have added a linear ramp to the blue of the shortest wavelength (630 nm) reported by Liebert et al. We assume for simplicity that the blueward ramp decreases to zero flux at 400 nm.

At wavelengths longer than 1000 nm, we supplement the fluxes measured by Liebert et al in two steps.

First, we note that Liebert et al assign a spectral type of M9.5V to this star, but in a more recent paper (Gagliuffi et al 2014), a spectral type of M9.7V is also listed. Gagliuffi et al present infrared spectra for some of the stars in their sample. Unfortunately, J0149 is not included in the spectra which they illustrate in their paper. However, in their Fig. 4, they include the spectrum of an M9.9V star over a range of wavelengths extending from 0.9 to 2.4 µm. Among the sample of stars reported by Gagliuffi et al, this M9.9V star is the closest in spectral type to the M9.5V and M9.7V types listed for J0149. We therefore consider it plausible to use the 0.9-2.4 µm spectrum of the M9.9V star as a proxy for J0149. In the wavelength range from 0.9 to 1.0 µm, the Gagliuffi et al spectrum overlaps with that of Liebert et al (1999): using this overlap, we merge the spectra to obtain a single empirical curve from 0.63 µm to 2.4 µm in the units used by Liebert et al, i.e. $10^{-15}$ W m$^{-2}$ µm$^{-1}$.

Second, in order to estimate the contribution to the integrated flux of wavelengths longer than 2.4 µm, we assume a Rayleigh-Jeans irradiance $I(\lambda) = K/\lambda^4$ where K is chosen to match the observed irradiance I(2.4) at λ = 2.4 µm. With this choice, the contribution to the integrated flux at λ > 2.4 µm is equal to 0.8I(2.4).

The integrated flux longward of 400 nm is then normalized to have a value of 1370 W m$^{-2}$ and is plotted in Fig. 2 along with the solar data. Inspection of Fig. 2 shows that the cross-over for J0149 and the Sun occurs at a wavelength of 1.0 μm. We adopt this cross-over wavelength in our study of PS effectiveness in a planet in the HZ of J0149.

### 2.4. Quiescent and flaring spectra for YZ CMi

In order to take advantage of the cross-over wavelength at 800 nm, we need to have access to spectra of YZ CMi which extend to at least 800 nm. Moreover, to make comparisons between PS effectiveness during flares and in quiescent conditions, we need to have spectra extending to 800 nm for both the quiescent star and also during a flare.

In an extensive study of flare spectra, Kowalski (2012) included some 60 spectra obtained at various phases during flares in a sample of six flare stars (including YZ CMi), with spectral types ranging from dM3e to dM4.5e. However, as regards the wavelength scale, among the 60 (or so) spectra, some 40-45 extend in wavelength only to (about) 550 nm, and 4-5 others extend to only 680 nm. As regards spectra which extend as far as 800 nm (or beyond), we find four flare spectra (in Kowalski's Fig. 6.32) which extend from 350 to 920 nm: however, the calibrations of these 4 spectra are said to be suspect longward of 750 nm, and also around 550 nm. Two reliable spectra are reported extending from 340 to 920 nm, but they were obtained during the decay phase of two flares, rather than at the peak phase. Only one spectrum (in Kowalski's Fig. 6.31) obtained at a flare peak extends over the wavelength range we seek: this is for a flare labelled impulsive flare #3 (if3) on YZ CMi, and the spectra extends out to 920 nm.

As regards the star in quiescence, Kowalski (2012) provides (in his Fig. 6.31) the **irradiance I$_q$(λ)** of YZ CMi **in its quiescent state**: this spectrum is suitable for our purposes because it also extends to wavelengths of 920 nm. We have scanned the Kowalski quiescent spectrum and plot the scan in Fig. 3 (dotted curve: labelled YZ-q). In the plot, the ordinate is in the units of irradiance used by Kowalski (2012), i.e. 10$^{-12}$ W m$^{-2}$ μm$^{-1}$**.** (Note that these units are 1000 times larger than those used by Liebert et al (1999): the difference arises in part from the fact that J0149 lies at a greater distance than YZ CMi, and in part from the later spectral type of J0149.) The wavelength range that is plotted along the abscissa is chosen to be wide enough to allow the reader to see the cross-over wavelength at 800 nm: however, **when it comes time for us to quantify the effectiveness of PS (in Section 3.3 below),** we shall not actually make use of the spectrum longward of 700 nm.

Also in Fig. 3, we show Kowalski's **data for the irradiance** spectrum **I$_f$(λ)** at the time of flare peak (open squares connected by dotted line: labelled YZ-f). The units along the ordinate are the same as those used for the quiescent spectrum. To avoid confusion in Fig. 3 at the longest and shortest wavelengths, the wavelength range we use for flare radiation extends only from 400 to 700 nm. In Kowalski's notation, the curve YZ-f refers to radiation emitted by the flare alone, i.e. *excluding* the quiescent radiation.

**It is important to recognize that, based on observations of solar flares, where imaging can be performed, only one active region is typically involved in any particular flare, while the remainder of the sun's surface remains in its non-flaring (quiescent) state. By analogy, it seems likely that an active region on a flare star also occupies only a fraction of the surface of the star at the time of flaring.**

Therefore, during a flare, the flaring star actually emits a spectrum with an irradiance $I_{tot}(\lambda)$ which is a combination of the irradiance from the quiescent surface $I_q(\lambda)$ and the irradiance $I_f(\lambda)$ from the portion of the stellar disk which is undergoing flaring. We already know what $I_q(\lambda)$ is: it is provided in Kowalski's Fig. 6.31 in units of $10^{-12}$ W m$^{-2}$ µm$^{-1}$. If we were interested in determining the irradiance of the flaring region in absolute terms, then we would need to know an extra piece of information, namely, the fractional area of the stellar disk which is occupied by the flaring region. This fractional area can be expressed as a filling factor, ff. The irradiance we measure on Earth during a flare is determined by integrating over the visible disk of the flaring star such that the value of $I_{tot}(\lambda)$ is given formally by the weighted sum $I_{tot}(\lambda) = ff*I_f(\lambda) + (1-ff)*I_q(\lambda)$. Typically, the value of ff for any particular flare on an M dwarf star (where no imaging of the surface is available) is unknown. Therefore in general, although we can determine $I_q(\lambda)$ reliably (by restricting observations to time intervals which include no flares), it is difficult (if not impossible) to determine $I_f(\lambda)$ reliably in absolute units.

But what the observations do tell us reliably, is that the intensity of the overall output of radiant energy from the M dwarf, as seen on Earth, increases during a flare by a certain amount X. Moreover, the observations indicate reliably that the value of X varies systematically with wavelength: $X(\lambda) < 1$ at long wavelengths, and $X(\lambda) > 1$ (sometimes by as much as 10-100) in the blue. A formula which represents the numerical value of $X(\lambda)$ at any wavelength $\lambda$ can be expressed formally as the weighted sum of $ff*I_f(\lambda) + (1-ff)I_q(\lambda)$ divided by $I_q(\lambda)$. Thus, a measurement of $X(\lambda)$ at any wavelength $\lambda$ already *incorporates the effects of the filling factor* on the overall output of radiant energy from the flaring M dwarf even if we cannot separate out the value of ff. If, e.g. $X(\lambda)$ is observed in a particular flare to have a value of 3, this could be due to a flare which occupies 10 percent of the surface (i.e. ff = 0.1) which has an enhancement $I_f(\lambda)/I_q(\lambda)$ of order 30, or it could be due to a flare which occupies 1% of the surface (i.e. ff = 0.01) which has an enhancement $I_f(\lambda)/I_q(\lambda)$ of order 300. Or it could be due to any other combination of flare intensity and flare area such that the overall effect is to increase the brightness of the star by a factor of 3. We have in general no way of knowing which of these options are relevant to any particular flare. But what we do know is that during the flare, the radiant energy received by a PS plant located on any HZ planet around the M dwarf will increase by a factor of 3.

There is one case in which we would really need to know the value of $I_f(\lambda)$ in order to proceed with the calculation in the present paper. Suppose we were dealing with a planet which belongs to the class known as "hot Jupiters". Such a planet is in an orbit that lies very close to the parent star, within perhaps only a few stellar radii above the surface. If this planet were to pass over the flare site at a distance of close to one stellar radius, then the flare site, as viewed by the planet, could essentially fill the entire field of view of the stellar surface all the way to the horizon. In such a case, PS plants on that planet could certainly be exposed to the full effects of the flare radiation: then the enhancement in radiant energy impacting the planet during a flare could indeed be as large as the (illustrative) factors of 30-300 mentioned above. However, in such cases, if the planet were really in such a close orbit, it would be too hot to lie in the HZ: such a planet would be of no interest as regards the study of living organisms (such as those which cause PS). Instead, we are interested here in planets which lie in the actual HZ of the M dwarf: in such cases, the separations between star and planet in terms of stellar radii are large enough (several dozen radii) that the flare radiation will typically *not* fill the entire field of view all the way to the horizon. Instead, the field of view from the planet will include not merely the flare, but also the portions of the star's surface which are not flaring at all. Therefore, the radiant energy arriving at the planet will not consist of flare radiation alone, but of flare plus

**quiescent radiations. As a result, the radiation arriving at the planet will be enhanced during a flare by the integrated effects of quiescent radiation (covering most of the stellar surface) plus the effects of the flare (with enhancements of (say) 30-300) diminished by the filling factor. In effect, a planet in the HZ will see a similar relative enhancement in radiation during a flare as an observer on Earth sees.**

**In view of this, and especially since we are interested in performing a *differential* study of PS effectiveness during flares on a particular star, we consider it reasonable to apply the empirical values of X(λ) as measured on Earth during a flare to be applicable also to the increase in radiant energy experienced by a plant on a planet which is in HZ orbit around that particular star I.e. at a radial distance of 80 stellar radii in the case of AD Leo. We note also that in the case of the cool star Trappist 1 (with radius 0.121 R(sun)), the planet labelled planet f in the HZ occurs at a distance of 0.039 AU, i.e. at a distance of about 70 stellar radii. Thus, planets in the HZ around an M dwarf orbit at distances which are many tens of stellar radii. These distance are large enough that the plants on such a planet will receive radiation not merely from the flare site, but in fact mostly from the non-flaring regions of the star: plants on such a planet will indeed be subject to radiant energy from the flare site, but (unlike the "hot Jupiter" case) the full effects of the flare site will be diluted by the factor ff.**

As regards PS effectiveness, we note that prominent in the flare spectrum is a black body (BB) continuum rising strongly towards the blue: this BB radiation presumably arises from dense gas in the photosphere which is locally heated by transient energy input (either particles or photons) coming down from the site of the primary flare energy release. (In solar flares, primary energy release occurs in the corona at altitudes of tens of thousands of kilometers above the photosphere: cf. Aschwanden et al 1996.) The temperature of the BB which provide the best fit to the continuum in the sample of flares reported by Kowalski (2012) ranges from **as low as** 7 thousand **to as large as** 14 thousand K: **we shall use both of these limits on the BB temperature when we model the spectra of flares for which we have no direct observations over a range of wavelengths where BB radiation is known to dominate**. Kowalski also reports that flares exhibit a Balmer continuum: however, this is not pertinent in the present paper because the photons lie at wavelengths <360 nm) which are outside the range (400-700 nm) which is relevant for PS.

Also prominent in flare spectra are four prominent lines of the Balmer series (Hα – Hδ) in emission at 656, 486, 434, and 410 nm: these emission lines also make a contribution to enhancing PS effectiveness during flares. According to Kowalski (2012: his Table 8), the Hα line alone can contribute as much as 10% of the overall flare flux in certain gradual flares. (In impulsive flares, the contribution is closer to 1%.) Also according to Kowalski (his Table 8), the fluxes in each of the three higher Balmer lines are equal to the Hα flux within a factor of 2, sometimes exceeding Hα, and in other flares being weaker.

**Our goal in this paper is to quantify how flares can be beneficial for life (specifically, as regards PS). But we also need to admit that not all flare phenomena are beneficial: in Section 6, we will discuss how UV photons and energetic particles generated by a flare may negatively impact life**.

*2.5. Normalizing the Sun and YZ CMi to Evaluate Relative PS Effectiveness*

In the units used by Kowalski (2012), i.e. $10^{-12}$ W m$^{-2}$ μm$^{-1}$, our inspection of the YZ-q spectrum provided by Kowalski (2012) indicates that the flux at the cross-over wavelength (800 nm) is 6.8. We therefore use this number to normalize the solar spectrum in order to perform a differential study between PS on Earth and on a planet located in the HZ of YZ CMi. That is, we change the solar irradiance at 800 nm from the value reported by Neckel and Labs (1140 W m$^{-2}$ μm$^{-1}$) to the value reported by Kowalski for YZ CMi (6.8 X 10$^{-12}$ W m$^{-2}$ μm$^{-1}$). In Fig. 3, the uppermost curve shows the solar spectrum scaled to the value 6.8 x 10$^{-12}$ W m$^{-2}$ μm$^{-1}$ at 800 nm.

## 2.6. The Quiescent and Flaring Spectrum of a Late M Flare Star

Liebert et al (1999) obtained spectra for J0149 in quiescence and during a flare spanning the wavelength range from 630 nm to 1000 nm. These spectra are useful for our purposes because they include the cross-over wavelength (1000 nm) for an M9.5-M9.7 dwarf. This enables us to normalize the solar spectrum so that we can make meaningful comparisons of PS effectiveness.

For PS purposes, we need to extrapolate the J0149 spectra reported by Liebert et al towards shorter wavelengths. In the case of the quiescent spectrum, Liebert et al (1999) measured the fluxes down to wavelengths no shorter than 630 nm (see the bottom panel in their Fig. 2). Our choice of linear ramp between 630 nm and 400 nm (see Section 2.3) means that we will miss some of the (small) features around 580 and 540 nm, as well as the higher Balmer lines (Hβ – Hδ). However, the latter lines are expected to be weak, especially since even the Hα line (expected to be the strongest in the series) can barely be identified in the quiescent spectrum reported by Liebert et al. In Figure 4, the resulting quiescent spectrum which we adopt for J0149 is shown as the lowest lying curve: the ordinates are in units of 10$^{-15}$ W m$^{-2}$ μm$^{-1}$.

In the case of the flare spectrum of J0149, we use the plot given by Liebert et al (their Fig. 3, top panel) for radiation at wavelengths >630 nm. Our scan of this spectrum appears in our Fig. 4 (dashed curve with crosses). These data include an exceptionally strong Hα line, with an equivalent width of 300 Å: using a scaling between Hα and total flare output, Liebert et al report that the flare output during the impulsive phase (lasting 1-2 thousand seconds) may have exceeded the bolometric photospheric luminosity.

In order to fill in the remainder of the PS-effective flare photons at wavelengths <630 nm, we impose a BB continuum (see Section 2.4 above). In view of the results reported by Kowalski (2012), a temperature **in the range T = (0.7-1.4) x 10$^4$ K can be assigned to the BB in different flares. In reporting our results below, we shall start with a model where we use T=10,000 K. However, we shall also report on results for a flare in which the BB temperature is only 7,000 K, and also results for a flare where T = 14,000 K**. We used the on-line tool SpectralCalc.com to calculate the spectral irradiance from a 10$^4$ K BB at wavelengths between 400 and 700 nm, and we fitted the output to the shortest wavelength flare point (630 nm) reported by Liebert et al. The resulting flare spectrum is illustrated in Fig. 4 (middle curve). **As Fig. 4 illustrates**, **in going from a wavelength of 630 nm to a wavelength of 400 nm, the BB curve increases from the measured irradiance at 630 nm (2.49 in the units of Fig. 4) to a value of 5.97 (in the same units). This increase in irradiance at 400 nm relative to that at 630 nm (by a factor of ρ = 2.4) is the increase reported by SpectralCalc.com for T=10,000 K. Although, for purposes of avoiding confusion, we do not include in Fig. 4 the BB continua which we will use for the cases T = 7 and 14 thousand K, we can report that at T = 7,000 K, the irradiance at 400 nm is larger than at 630 nm by a factor ρ = 1.44. At T = 14,000 K, the corresponding factor is ρ = 3.3. Thus, in going from T = 10,000 to**

14,000 K, the value of ρ increases linearly with T. But at lower temperatures, from 10 to 7 thousand K, ρ falls off more steeply, varying as $T^{1.4}$. Because of the different values of ρ, we expect that in flares where the BB temperature is 14 (or 7) thousand K, PS will be more (or less) effective than the results for T = 10 thousand K. We shall report results for all 3 values of T below.

Referring to Figure 4, we see that there are no emission lines of Hβ – Hδ on the curve between 630 and 400 nm: only a continuum is included, although the Balmer lines can be prominent in flares. Therefore, the PS effectiveness we will derive for the "flare" in Fig. 4 will provide only a lower limit on the actual value. To avoid confusion with the quiescent spectrum in Fig 4, we plot the flare spectrum only over the range of wavelengths from 400 to 700 nm.

### 2.7. Normalizing the Sun and J0149 to Evaluate Relative PS Effectiveness

In units of $10^{-15}$ W m$^{-2}$ μm$^{-1}$, our inspection of the quiescent spectrum in Liebert et al (1999) indicates that the flux at the cross-over wavelength (1000 nm) is 10.8. As we did for YZ CMi, we can set the stage to perform a differential study of PS effectiveness between the Sun and J0149 by changing the solar radiative flux at 1000 nm from the units used by Neckel and Labs (760 W m$^{-2}$ μm$^{-1}$) to the units used for the J0149 radiation (10.8 x $10^{-15}$ W m$^{-2}$ μm$^{-1}$).

In Fig. 4, the uppermost curve shows the solar spectrum scaled to J0149 at 1000 nm.

### 3. Oxygenic Photosynthesis

The only chemical compound which we consider in this paper as responsible for PS is chlorophyll. We recognize that this sets a limit on the processes we are considering. Nature has developed various organisms (phototrophs) that can capture photons in order to perform certain functions. E.g. retinal is the basis for animal vision, and may assist in certain organisms to convert light into metabolic energy. Moreover, some phototrophs, e.g. haloarchaea, can absorb photon energy in order to create ATP, and can move chloride ions in order to generate a voltage gradient to energize reactions. Another subset of phototrophs, those involved in anoxygenic PS (e.g. purple bacteria), cause the release of sulfur from $H_2S$. However, none of these photo-sensitive reactions involve fixing carbon from its inorganic form ($CO_2$).

In the one biosphere where we know that life has been successful, the most common phototrophs participate in oxygenic PS. This participation involves the use of photons absorbed by chlorophyll to perform a number of tasks: (i) convert $CO_2$ into organic material, (ii) split water so as to donate electrons to an electron transport chain, and (iii) cause $O_2$ to be released into the environment. The combination of these tasks is what makes us decide to concentrate in the present paper on oxygenic PS in the context of extraterrestrial life.

In order to quantify the effectiveness of oxygenic PS in the presence of various radiation environments, we need to have specific information about the wavelength-dependences of the absorbance of various kinds of chlorophyll. We now turn to a consideration of this information.

### 3.1. Chlorophyll components in plants

Chl occurs in various forms. Chl-a and Chl- b are the commonest chemical variants which occur in wild plants on Earth, with varying ratios depending on the species. Chl-a/b ratios range from 3.3 to 4.2 in species which grow in sunlight, but the ratio falls to 2.2 in plants which grow in shade (Munns et al 2016). Shade-adapted plants adapt to the lower light levels by containing a larger fraction of Chl-b: this allows the plants to extend the range of wavelengths over which light is effectively absorbed towards the blue. Organisms on Earth cannot simultaneously maintain effective ability to harvest light when the level of light intensity is high, and when the light intensity is low. When the light becomes more intense, the PS capacity must be ramped up so that the plant can acclimatize to the higher light level: this ramping up requires a finite time $T_r$. In the context of a planet in orbit around a flaring M dwarf, the relevance of the calculation we perform in this paper will depend on how the ramping time $T_r$ compares to the duration of an individual flare $T_f$. Values of $T_f$ span a broad range: for details, see Section 5(i) below. In the limit of short flares, i.e. when $T_f \ll T_r$, there will not be enough time for ramping up to acclimatize to the higher light level during a flare. In such cases, the PS system may be overloaded with excess light, and it is unlikely that any significant enhancement in PS efficiency will occur during such flares. However, for long flares, i.e. $T_f \approx T_r$ or greater, the PS system will have time to acclimatize to the higher light level, and the process we report on here could be effective. For present purposes, it is important to note that observers report a correlation between flare amplitude and flare duration: the more intense the flare, the longer it lasts (e.g. Kowalski et al 2013). Therefore, the shortest flares (i.e. those with $T_f \ll T_r$) tend to be of weak intensity, and are therefore not expected to contribute significantly to the effects we discuss here. The largest effects will be contributed by the flares with largest amplitude: these are also the flares which last longest, and are therefore more likely to satisfy the condition $T_f \geq T_r$. In such flares, acclimatization to the higher light level will have time to occur.

**In view of the above discussion, it must be admitted that not all flares on an M dwarf are necessarily favorable for PS. The flares which are optimal for PS are the longer-lived flares. In Section 5 below, we present detailed information on how long the individual flares last, and how frequently the longer-lived flares are observed to occur.**

### 3.2. Chlorophyll absorbances as a function of wavelength

As regards empirical studies of Chl absorbances, we note that the wavelength behavior depends on the solvent, and on how the solution was prepared, the initial Chl absorption curve reported by Zscheile and Comar (1941) was obtained by using ether as the solvent: the results identified peaks in absorption at 429 and 659 nm for chl-a, and at 455 and 642 nm for chl-b. The peaks were rather sharp, with FWHM of 20-30 nm. Thus, both variants pf Chl show two absorption peaks, one in the red, one in the blue, with a broad range of low absorptions from (roughly) 490 to 630 nm. In a living organism, where the Chl exists in a medium containing a mixture of proteins and pigments, the peaks in absorption are still present, but they are shifted and broadened.

In order to proceed with a quantitative discussion, we note that the wavelength-dependent absorbances of Chl exhibit slightly different behavior depending on the environment in which the Chl is located. In the present paper, two different environments are considered (see Munns et al 2016, their Section 1.2.2). First, light-harvesting Chl (LHC) protein complexes: in these, light is collected by many pigment molecules located in the photosynthetic membrane. Second, the reaction centers (RC): these are situated at the center of the LHC, and they act as antennas for the initial (PS II) stage of the PS process. Because there are (at least) two variants of Chl that are used in plants, we need to consider in what ratios they are present in order to evaluate PS efficiency.

In this regard, we note that both Chl-a and Chl-b are bound to the LHCs. The availability of Chl-b Is important for assembly of proteins in the LHC complexes in plants ([Bellemare et al., 1982](); [Peter and Thornber, 1991]()). The presence of Chl-b in plants plays an important role in stabilizing the LHC complex. If Chl-b is absent, the LHC complex becomes unstable, leading to formation of proteases and protein degradation: thus, although Chl-b is in the minority in plants, that does not mean that we can assume an abundance of zero.  In the context of stellar flares, it is important to note that the amount of Chl-b depends on the intensity of the incoming light: plants use lesser amounts of Chl-b when the light intensity is high, but they use more when the light intensity is low. This suggests that, when an M dwarf has a flare in progress, plants on a nearby planet might in principle reduce their Chl-b content. However, this principle should not be pushed too far: the LHC will still need *some* Chl-b in order to stabilize the complex under varying light intensity. (E.g. in a laboratory study of extinction coefficients [Evans and Anderson 1987], the ratio of Chl-b to Chl-a was allowed to fall to 10-12%, but no lower.) Moreover, plants are known to acclimatize chloroplast movement during varying light intensity: in the stronger light intensity which arrives during a flare, e.g. the chloroplasts will have a tendency to move farther into the interior of the plant. It is also important to recognize that other pigments such as flavonoids and betalains may also play a role in light harvesting under varying light intensity. Therefore, it is expected that plants on a planet that is illuminated by an M dwarf will have the capability to undergo certain trade-offs as the plants strive to adjust to conditions of varying light intensity. In view of these various effects, we speculate that plants might in favorable circumstances develop somewhat different PS absorbance spectra during high intensity flares, i.e. the absorbance spectra of photosynthetic organisms may not be the same in planets around other stars. The absorbances may become adapted, and may even become time-dependent, in order to adapt to both the spectrum and intensity of the parent star. However, if we wished to treat this problem in full detail, we would need to know the time-scales $T_r$ on which the various adjustments occur. Such a task is beyond the scope of the present paper, where we are interested in a proof of principle.

### 3.3. The essential calculation in this paper: convolution

To quantify and compare PS effectiveness (PSE) in different environments, our approach will be to convolve the wavelength-dependent photon spectra (in Figs. 3 and 4) with the wavelength-dependent absorbance (in Fig. 5).

In the present calculation, we will use the wavelength dependences of the absorbances in LHC and RC environments as shown in Fig. **5** over the range of wavelengths from 400 to 700 nm. The curves in the figure are taken from Munns et al (2016): the figure shows *in situ* absorption spectra (eluted from gel slices) for pigment-protein complexes corresponding to photosystem II reaction centers (PSII RC) and light-harvesting Chl (a,b)-protein complexes (LHC). A secondary peak at 472 nm, and a shoulder at 653 nm, indicate contributions from Chl b to the absorption spectra. The units along the ordinate axis are dimensionless, but they can be converted to the absolute extinction coefficient EC by noting the following calibrations for the red peaks when in a particular solvent (Munns et al 2016): for Chl-a at 665 nm (with absorbance ≈ 0.51 in Fig.1), EC = 71.4 L mmol$^{-1}$ cm$^{-1}$, while for Chl-b at 653 nm (with absorbance ≈ 0.27 in Fig. 1), EC = 38.6 L mmol$^{-1}$ cm$^{-1}$. (Note that the ratio of dimensionless absorbances Chl-a/b is 1.89, i.e. essentially identical to the ratio of the absolute values of EC: 1.85.) However, we do not need to know the absolute absorbances in the present paper: the availability of relative absorbances (as shown in Fig. 1) is sufficient to enable us to do a differential study between various stars.

As well as the two peaks which dominate the curves in Fig **5**, it is noteworthy that the absorbance has a relatively sharp long-wave limit at 680-700 nm. The energy of photons with such wavelengths is 1.77-1.82 eV: these are lower limits on the photon energy that is required for oxygenic PS to occur. The limits are associated with the energy gaps between ground state and excited electronic levels in the chlorophylls which occur in the RC's of PSII (P680) and PS I (P700). Slight relaxations of the lower limits on energy can occur in the presence of Chl d, where the longest absorption wavelength is 710-715 nm (Miyashita et al 1997), and in Chl f, where the corresponding figure is 706 nm (Chen et al 2010). For the purpose of evaluating oxygenic PS effectiveness, we will not make significant errors by limiting our consideration to photons which have wavelengths of 700 nm and shorter.

In the convolution study that follows, we scan the absorbance curves in Fig. 5 and set up arrays of spectral absorbances for light harvesting complexes, ALHC(i), and absorbances for reaction centers, ARC(i). Here, the index i denotes a set of n = 301 wavelengths between λ(1) = 400 and λ(301) = 700 nm.

To start our calculation, we first obtain an array R(i) of irradiances at the same set of n wavelengths for each spectrum in Fig. 3 and in Fig. 4 in the range 400-700 nm by interpolating in wavelength according to the same recipe, i.e. each spectrum is interpolated at an array of 301 wavelengths spanning the range from 400 nm to 700 nm in steps of 1 nm.

Then if it were true that PS depended on the *flux of radiant energy*, we could obtain a measure of the PS effectiveness (PSE) by convolving R(i) with the absorbance at each wavelength. However, PS does not depend on the flux of *energy*: instead, PS is a quantum process that depends on the *flux of photons*. To allow for this effect, we need to convert the irradiance value at each wavelength to a quantity that is proportional to the flux of photons at that wavelength. Since each photon has an energy E(ph) = hf = hc/λ, we could in principle do this conversion by dividing R(i) at each wavelength by the corresponding E(ph). Equivalently, we can obtain a quantity that is proportional to the flux of photons by multiplying R(i) by λ(i). In what follows, for convenience we choose to express the values of λ(i) in units of microns. We therefore can obtain a quantity which is proportional to PSE for the two distinct sites LHC and RC as follows:

PSE(LHC) = Σ(i=1,n) R(i) * λ(i) * ALHC(i)                     (1)

PSE(RC) = Σ(i=1,n) R(i) * λ(i) * ARC(i)                        (2)

Although we have selected for definitiveness the absorbances in Fig. 5 to illustrate quantitative details, this is not meant to imply that other possible forms of PS are excluded from occurring on planets around other stars. E.g. in a hydrogen dominated atmosphere, where the principal carbon species may be $CH_4$, PS based on methane might be the main source of energy for life (Bains et al 2014). The effect of flare photons on methane-based PS is not discussed in this paper. Or sulfur-based photosynthesis might be the dominant energy source in the vicinity of deep-sea thermal vents (black smokers), where the photons are associated with the thermal vent, where the temperature is only 600 K (Raven and Donnelly 2013). It is unlikely that photons from flares (such as we discuss here) would have any significant effect on processes at the bottom of the ocean.

In the present paper, we confine attention to oxygenic PS, and we restrict attention to the wavelength range 400-700 nm.

## 4. Quantitative Comparison of Photosynthetic Effectiveness (PSE) on Earth with PSE on Planets in the HZ of Flare Stars

Now that we have similar arrays for the various spectra, we can make meaningful comparisons of PS effectiveness (PSE) by convolving an R(i) array with λ(i) and also with either the ALHC array or the ARC array, in accordance with eqs. (1) or (2) above.

### 4.1. Comparing PSE on Earth and on a planet of YZ CMi

To start the comparisons, we use the YZ-q normalization of the solar spectrum as described in Section 2.5 above. In combination with eq. (1), we can obtain the convolution of R(i)* λ(i )*ALHC(i) which is appropriate for conditions on planet Earth. We find that the numerical value of PSE(LHC) is 391 (in units of $10^{-12}$ W m$^{-2}$ μm$^{-1}$). For RC absorbance, the convolution in eq. (2) leads to a numerical value of PSE(RC) is 389 (in the same units). These numerical values are a measure of the effectiveness of PS on Earth in units which allow us to make a meaningful comparison with conditions on a planet which is in the HZ of YZ CMi. The numerical values are listed on the first row in Table 1, with the notation (§2.5) referring to the normalization of the solar spectrum described in Section 2.5 above.

Next, we use the R(i) spectrum for YZ CMi in its quiescent state, and calculate the PSE for this spectrum. Results for the same measure of PSE under these conditions are listed in row 2 of Table 1. We see that, when YZ CMi is in its quiescent state, the PSE on a planet in the HZ of YZ CMi is about one order of magnitude smaller than on Earth. Then, we use the R(i) spectrum of YZ CMi in its flaring state. Results for the same measure of PSE under these conditions are listed in row 3 of Table 1. Actually, during a flare, the light from the star is the sum of the flare spectrum plus quiescent. Therefore, in row 4 of Table 1, we give the PSE values in the presence of the combined f+q photons. We see that during a flare, the PSE that is some 5 times larger than in quiescence. When YZ CMi is in a flaring state, the PSE on a planet in its HZ reaches values of 50-60% of the value of PSE on Earth. Thus, on a planet in the HZ of YZ CMi, flares on the parent star lead to PSE's on the planet which are within a factor of 2 of PSE's on Earth.

TABLE 1. Quantitative estimates of a measure of photosynthetic effectiveness (PSE) for different stars and different PS sites

| Star | PSE(LHC) | PSE(RC) | PSE(Earth/star) |
|---|---|---|---|
| Sun/YZ (§2.5) | 391 | 389 | 1 |
| YZ CMi (quies) | 43 | 43 | 0.11 |
| YZ CMi (flare) | 173 | 179 | 0.45 |
| YZ CMi (f+q) | 216 | 222 | 0.56 |
| Sun/J0149 (§2.7) | 944 | 940 | 1 |
| J0149 (quies.) | 7.3 | 7.7 | 0.008 |
| J0149 (flare) | 151 | 156 | 0.16 |
| J0149 (f+q) | 158 | 164 | 0.17 |

*4.2. Comparing PSE on Earth and on a planet of J0149*

To start the comparisons, we use the J0149 normalization of the solar spectrum as described in Section 2.7 above. Then we repeat the steps described in Section 4.1. For the J0149 normalization, we find that the numerical value of PSE(LHC) on Earth is 944 (in units of $10^{-15}$ W m$^{-2}$ μm$^{-1}$). For RC absorbance, the convolution in eq. (2) leads to a numerical value of PSE(RC) on Earth of 940 (in the same units). These numerical values are a measure of the effectiveness of PS on Earth in units which allow us to make a meaningful comparison with conditions on a planet which is in the HZ of J0149. The numerical values for the Earth are listed in row 5 in Table 1, with the notation §2.7 referring to the normalization of the solar spectrum described in Section 2.7 above.

Next, we use the R(i) spectrum for J0149 in its quiescent state, and calculate the PSE for this spectrum. Results for the same measure of PSE under these conditions are listed in row 6 of Table 1. We see that PSE is more than 2 orders of magnitude lower than on Earth.

Lastly, we use the R(i) spectrum of J0149 in its flaring state. **In the first instance, we assume that the flare radiation contains a BB component with a temperature of T = 10,000 K.** Results for our measure of PSE under these conditions are listed in row 7 of Table 1: the numerical results are 151 and 156, i.e. some 20 times larger than in quiescence. However, as was noted above **(in Section 2.6), in a flare where the BB temperature is 14,000 K, the PSE effectiveness is expected to be larger: in fact we find the values to be 179-186, i.e. larger by some 19% than the results for T = 10,000 K. On the other hand, in a flare where the BB temperature is only 7,000 K, the PSE effectiveness is found to be 118-120, i.e. reduced by some 30% relative to the results for T = 10,000 K. Thus, when we allow for the known**

**range of BB temperatures in flares, photons from flares in J0149 are more effective than quiescent photons as regards PS by factors of 16-24.** When J0149 is in a flaring state (including the contribution of quiescent photons (see row 8 in Table 1), the PSE is **found to lie in the range 13-20 percent of the** value of PSE on Earth. Thus, although flares do cause an increase in PSE in this case, the PSE on a planet in the HZ of J0149 remains, even during a "spectacular flare" such as that reported by Liebert et al (1999), almost one order of magnitude smaller than on Earth.

5. Contributing Factors to PS Effectiveness: Flare Durations, Intensities, and Frequencies

Although a flare involves a significant increase in photon output from a flare star, the overall PS effectiveness on a planet in orbit round the flare stars is also going to depend on other factors as well. Here, we discuss three of these: (i) how long does a flare last? (ii) how intense is the radiation? and (iii) how often do the flares occur?

(i)As regards flare durations, Gershberg (2002) includes an extensive discussion of data which were collected by observers over a time period of several decades. These data indicate that the duration of a flare $T_f$ (defined as the time interval during which the star's brightness is greater than the quiescent level) span a range of several orders of magnitude. At the long end of the distribution, the longest recorded flare (on the active flare star AU Mic) lasted 30-40 hours, as observed in the far ultraviolet. A flare on an Orion flare star (T177) lasted about 20 hours, as recorded photographically. The star AZ Cnc was observed by the ROSAT X-ray satellite to have a duration of at least 8 hours. ROSAT also observed a flare on the star BY Dra which lasted about 4.5 hours. The star EV Lac was observed in visible light to have flares lasting 5 hours and 4.5 hours: the same star had an X-ray flare which lasted more than 4 hours. Coordinated ground-based observations of the star YZ CMi across a wide range of longitudes discovered an optical flare lasting more than 4 hours. A flare on the star EQ Peg was observed to have a duration in X-rays of more than 2.5 hours. Thus, the data indicate that flares on M dwarf stars can and do last as long as one hour and more. The longest flares also are associated with the largest flare energies, so they have the best opportunity to have an effect on the PSE.

Further data on flare durations have been reported by Kowalski et al (2013: their Fig. 18): in their sample of flares, which by no means includes the largest flares ever recorded, the maximum duration extends to time-scales of order 1 hour. In particular, the emission in Hα lasts longer than at any other visible wavelength: the half-life for Hα in some of their flares is of order 1 hour. Thus, the Hα radiation from a flare as a whole can last for at least 2 hours. This longevity of Hα photons from a flare is important for flare-related PS effectiveness, since the Hα line lies close to the red peak of the PS absorbance (see Fig. 1).

As regards the flare on J0149, Liebert et al (1999) remark that their data spanned a time of only 42 minutes, but they could not determine "what fraction of the total flare event was detected, or whether the observations included the strongest part of the flare". But their data indicated certainly that the flare lasted at least 42 minutes. And the data which they did have suggested that for a period of 1-2 thousand seconds, the flare output exceeded the total output power from the star as a whole.

At the extreme short end of the distribution of flare durations, Gershberg (2002) reports that some flares last for no more than 1 second.

As was mentioned in Section 3.1 above, the duration of a flare $T_f$ has a bearing on PSE. Specifically, if $T_f \ll T_r$ (the ramping-up time-scale for PS to adapt to higher light levels), then a plant will not be able to adapt to the higher light level in the flare. Certainly the shortest flares mentioned above (lasting only seconds or a few minutes) will fall into this category, and are of little relevance in the present context. On the other hand, the longest-lasting flares on M dwarfs (with durations of order one hour or more) may well last long enough to allow the terrestrial light harvesting apparatus to acclimatize to the higher light levels. If $T_r$ is of order (say) 1 hour, we speculate that the enhancements in PSE calculated above (see Table 1) pertain preferentially to longer-lasting flares, unless planets near M dwarfs have developed a method for their light harvesting apparatus to adapt to changes in light conditions which are more rapid than terrestrial organisms have to deal with.

(ii) As regards the intensity of radiation which is emitted by flares, we have already quantified this in Figures 3 and 4, where one can make a direct comparison between the intensity of radiation from a star in its quiescent state, and the intensity at the same wavelength in the flaring state. In Figures 3 and 4, we see that at the longest wavelengths of PS sensitivity, the ratio of flare radiation to quiescent radiation is close to unity. But at the shortest wavelengths of PS sensitivity (400 nm) the ratio is several dozen.

(iii) As regards the frequency of flares, Shakhovskaya (1989) has compiled data on how frequently flares of a given energy occur on each star. The energies plotted by Shakhovskaya are given in terms of $E_B$, the energy contained in the Johnson B band (spanning wavelengths from 360 to 550 nm): to convert these to total energies in visible photons, we need to multiply by 3.8 (Gershberg 2002). Inspection of Shakhovskaya's Fig. 2 suggests that in flare stars such as YY Gem or GT Peg, flares with $E_B$ of order $10^{26}$ J occur roughly once every 10 hours. The optical luminosity in the form of such flares is expected to be of order $3.8 \times 10^{26}/36000 = 10^{22}$ W. Adding in flares of other energies from as much as $10^{28}$ J (occurring rarely, once per 1000 hours) to as little as $10^{24}$ J (occurring once every hour or so), the integrated flare luminosity may be of order $L(f) = 3 \times 10^{22}$ J. How does this compare with the quiescent luminosity $L_{bol}$ of these stars? In Gershberg's (2002) list of 463 flare stars, $L_{bol}$ is listed for only 20% of the sample: these are the brightest and/or nearest stars, and their listed log $L_{bol}$ (in J) values range from 22.39 to 26.55, with a mean value of about 25.0 and an rms deviation of $\sigma = \pm 0.9$. For the remaining >80% of the stars in Gershberg's list for which no value of $L_{bol}$ is given, it is expected that these flare stars have log $L_{bol}$ < 25 by a finite amount. There is no indication as to what the finite amount may be, but as a rough guide, we estimate 1-2σ. If this is correct, then the great majority (>80%) of flare stars in Gershberg's list have a mean log $L_{bol}$ in the range 23-24, i.e. a mean $L_{bol}$ of order $3 \times 10^{23}$ J.

In view of this, the fraction of output power from flare stars in the form of flares. $L(f)/L_{bol}$ may rise to values as large as 10% (see Mullan 1989). Although the fractional flare output will be less than this limit in some flare stars, we suggest that there may be a selection effect in favor of life in the following sense: the effectiveness of flare-enhanced photosynthesis on a planet around a flare star will be greater in cases where the star has (a) longer flares and (b) outputs a larger fraction of its total power in the form of flares. In the context of the present paper, we suggest that such stars could be optimal candidates in a search for life around M dwarfs.

## 6. Discussion

Visual inspection of the flare spectra in Figs. 3 and 4 suggests why flares contribute to an increase in PS effectiveness: the light intensity increases across essentially all of the visible spectrum where the photons **are maximally effective in driving** PS. Especially important during a flare in this regard is the presence of a black-body continuum at temperatures of order $10^4$ K: this continuum provides an enhanced supply of photons which overlaps favorably with the absorbances of Chl (LHC) and Chl(RC). As well as the continuum, the presence of enhanced emission lines during a flare also contributes to enhancing PS. The most prominent line feature of a flare spectrum at longer wavelengths (>600 nm) is the H$\alpha$ line which is strongly in emission at 656 nm. Comparing with the Chl absorbances in Fig. 5, it is clear that H$\alpha$ has overlap with the red peak of Chl absorbance, especially with the LHC "shoulder". And at the shorter wavelengths (<500 nm), the higher members of the Balmer series (at 486, 434, and 410 nm), all of which are also strongly in emission, overlap with the blue peak of Chl absorbance, especially with the LHC data. **It is almost as if the optical spectra of M dwarf flares are (in a sense) "tuned" to exploit optimally the wavelength-dependent absorbances of Chl.**

We note that Ranjan et al (2017) have already demonstrated that photons emitted by stellar flares could be beneficial for pre-biotic chemistry. The difference between our work and that of Ranjan et al is that we focus on visible photons, whereas Ranjan et al focused on UV photons. But our overall conclusions have an interesting overlap, **specifically,** some flare photons may be beneficial for life.

Unfortunately, other aspects of flare activity may be detrimental to life. For example, extreme UV photons can be hazardous to biomolecules (Sagan, 1973). Moreover, coronal mass ejections (CME's) may strip away some of the atmosphere of an orbiting planet if the latter does not possess a significant magnetic field (Kay et al 2016). In the case of the Sun, CME's are known to be associated with certain flares: during the years from 1996 to 2010, the GOES spacecraft recorded 22,688 solar flares, while during the same period, the SOHO spacecraft recorded 12,433 solar CME's (Youssef 2012). That is, a CME occurs on average in about 50% of solar flares; however, the percentage of flares with CME's increases to essentially 100% in the largest flares (those which are classified as X-flares: see Andrews, 2003). Stellar flares are known to have total energies which exceed the largest solar X-flare energies by several orders of magnitude (Shakhovskaya 1989). An advantage of the largest stellar flares would be that they may have photon fluxes which enhance the PS effectiveness results by factors which are even larger than the factors we have presented here. But, on the other hand, CME's may also be more common in association with the largest stellar flares. Whether flares as a whole lead to a net benefit or a net deficit as regards biology remains to be seen.

## 7. Conclusion

We have found that a quantitative measure of the effectiveness of oxygenic photosynthesis (PS) in planets which are in orbit in the HZ around M dwarfs is smaller than on Earth by 1-2 orders of magnitude when the M dwarf is in quiescence. But during flares, the PS effectiveness increases by factors of ~5-20. These increases bring the PS effectiveness on planets in the HZ around mid-M dwarfs up to values which approach within factors of 2-3 those on Earth. But in late-M dwarfs, even during flares the PS effectiveness remains almost one order of magnitude smaller than in the case of the Earth. This suggests that biology on a planet in the HZ of a mid-M dwarf could benefit from flares, at least as far as PS effectiveness is concerned. However, flares also lead to certain negative effects (Sagan 1973), so **despite the results presented here, it might be argued that when all the relevant effects will have been quantified, flares might turn out *not* to have a net benefit for life. On the other hand, given the propensity for life to adapt to a variety of conditions, it is possible that life on a planet around an M dwarf, where UV and optical photon excesses are more or less continuous hazards, may develop some way to protect itself against the negative effects of flares.**

In the event that flares turn out to have a net benefit for life on a planet that is in orbit within the HZ of an M dwarf, **we consider that it could be worthwhile to pay attention to two different time-scales. For brevity, we refer to these as the "daily" time scale and the "yearly" time scale. As regards the "daily" time-scale, the axial rotations of HZ planets around M dwarfs are possibly tidally locked to the orbital period. In view of this locking, an anonymous referee has suggested that "a continuous illumination of one planetary hemisphere can be advantageous for PS-based life".**

**As regards the "yearly" time-scale, if the occurrence of flares turns out to have a net benefit for life, then we would like to suggest the following possibility**: the "yearly" cycle of growth and dormancy which is characteristic of many plants on Earth might be replaced by a different type of cycle. On Earth, the tilt of the rotational axis relative to the orbital axis results in a modulation of photon fluxes at a given latitude on the Earth which has a time-scale determined by the orbital period (1 year). However, the orbital periods of HZ planets around M dwarfs are much shorter than the orbital period of Earth: e.g. in the Trappist 1 system, the planets in the HZ (e, f, g) have periods of 6-12 days. Given the finite time-scales required for chemical reactions to occur at temperatures of 288 K (the mean global temperature: we do not examine here the possibilities of higher/lower temperatures at different latitudes), it is difficult to imagine how plants on planets in the HZ of Trappist 1 (where the temperatures are also about 288 K) could go through complete cycles of growth and dormancy on periods of only a few days. On the other hand, if photosynthesis (and by extension, perhaps other biological processes) is enhanced by radiation from flares on the parent star, then the period $P_a$ of an activity cycle of that star could provide a natural process of waxing and waning of PS activity on time-scales of order $P_a$.

What sort of values of $P_a$ might stellar activity cycles exhibit? A report referring to an F5 star which was observed asteroseismically by COROT suggested a cycle with a period of order 120 days (Garcia et al 2010). In another F star (KIC 3733735), Mathur et al (2014) reported a cycle period of at least 1400 days. However, in a subsequent paper (Regulo et al 2016), the same F star was reported to exhibit a seemingly cyclic behavior in the frequency of its asteroseismic modes: inspection of Fig. 10 in Regulo et al (2016) suggests a period of order 400-500 days. The largest survey of activity cycle periods to date is based on Kepler photometry: Reinhold et al (2017) have found long-term variations in more than 3000 Kepler stars which suggest the occurrence of activity cycles. They identify cycle periods in the range 0.5-6 years. Such values of $P_a$ augur well for the possibility that photochemical reactions which are effective on time-scales of a few months on Earth (i.e. during the growing season in the northern hemisphere) will also have time to complete their functions in the HZ environment around a flaring M dwarf.

To consider this more quantitatively, suppose that (as was mentioned in Section 5(iii)) 10% of a flare star's energy is emitted in the form of flares. Consider a planet in the HZ of such a star: by definition, the total incoming energy flux to that planet (integrated over all wavelengths) is comparable to what the

Earth receives from the Sun (i.e. the solar "constant" ≈ 1370 W m$^{-2}$). However, in the case of the flare star's planet, most of the incoming radiant energy occurs at wavelengths around 1000 nm, with greatly reduced fluxes at the wavelengths which are relevant for PS (400-700 nm). As a result, in the quiescent state, we have found that the PS effectiveness is 10-100 times less than the Earth's value (see Section 4 above). But in the course of a flare, the PS effectiveness increases to as much as 50-60% of the Earth's value, i.e. to a flux of order 700 W m$^{-2}$ . With flares contributing 10% of the star's output, this means that for 10% of the time, the flare star's planet could receive roughly 700 W m$^{-2}$ of input power that is PS effective. Averaging over time, this planet can expect to receive a flux of roughly 70 W m$^{-2}$ of PS-effective photons: this is less than Earth receives by a factor of about 20. As a result, a growing season which on Earth receives enough PS-effective photons to be complete in (say) 3 months, would require about 5 years on a flare-star planet. We consider it noteworthy that a 5-year interval falls within the range of activity cycle periods that have been reported among low-mass stars by Reinhold et al (2017).

In contrast to the pessimistic conclusion of Sagan (1973) regarding the negative effects of stellar flares on life, the present paper suggests that the opposite conclusion may be worth exploring: flares may turn out to play an essential role in driving photosynthesis on a planet in the HZ around an M dwarf.

**ACKNOWLEDGEMENTS**

DJM thanks Dr Sukrit Ranjan for bringing valuable papers to his attention.

Figure 1. Determination of the cross-over wavelength for the Sun and the mid-M star AD Leo. Continuous line: irradiance spectrum received on Earth (normalized to a distance of 1 AU from the Sun). The line is plotted using data given by Neckel and Labs (1984). Crosses: irradiance spectrum received by a planet in the HZ of the flare star AD Leo. The crosses are plotted using data given by Cohen (2017), normalized such that the AD Leo planet has a mean temperature of 288 K.

Figure 2. Determination of the cross-over wavelength for the Sun and the late-M star J0149. Continuous line: irradiance spectrum received on Earth (normalized to a distance of 1 AU from the Sun). The line is plotted using data given by Neckel and Labs (1984). Crosses: irradiance spectrum received by a planet in the HZ of the flare star J0149. The crosses are plotted using data given by Liebert et al (1999) supplemented at shorter and longer wavelengths as described in the text (Section 2.3). The crosses are normalized such that the J0149 planet has a mean temperature of 288 K.

Figure 3. Spectra of the Sun and of a flare star with mid-M spectral type (YZ CMi). Abscissa: wavelengths in nanometers. Ordinate: energy flux in units of $10^{-12}$ W m$^{-2}$ μm$^{-1}$. Continuous curve: spectrum of the Sun. Original solar data were taken from Neckel and Labs (1984), and then normalized to equal the observed flux from YZ CMi in quiescence at a wavelength of 800 nm. Dotted curve: the spectrum of YZ CMi in quiescence. Dotted curve with open squares: the spectrum of YZ CMi at the peak of a flare. The spectra of YZ CMi are taken from Kolwalski (2012): the flare spectrum in our figure refers to the particular flare labelled if3 by Kowalski.

Figure 4. Spectra of the Sun and a flare star with late-M spectral type (J0149). Abscissa: wavelengths in nanometers. Ordinate: energy flux in units of $10^{-15}$ W m$^{-2}$ μm$^{-1}$. Continuous curve: spectrum of the Sun. Original solar data were taken from Neckel and Labs (1984), and then normalized to equal the observed flux from J0149 in quiescence at a wavelength of 1000 nm. Dotted curve: the spectrum of J0149 in quiescence. Dotted curve with crosses: the spectrum of J0149 during a flare. The spectra of J0149 are taken from Liebert et al (1999).

Figure 5. Absorbances of chlorophyll molecules (including Chl-a and Chl-b) as a function of wavelength in two different environments in plants. Line: photosystem II reaction center (RC). Crosses: light-harvesting complex (LHC). The absorbances are given in dimensionless units: for conversion to extinction coefficient in absolute units (Liters per millimol per cm), see Section 3.3. The wavelengths are in units of nanometers: the photons which we use for calculating photosynthetic efficiency (PSE) extend over a range of wavelengths from 400 nm to 700 nm. The absorbance curves have been sampled at intervals of 1 nm for our purposes in the present paper.

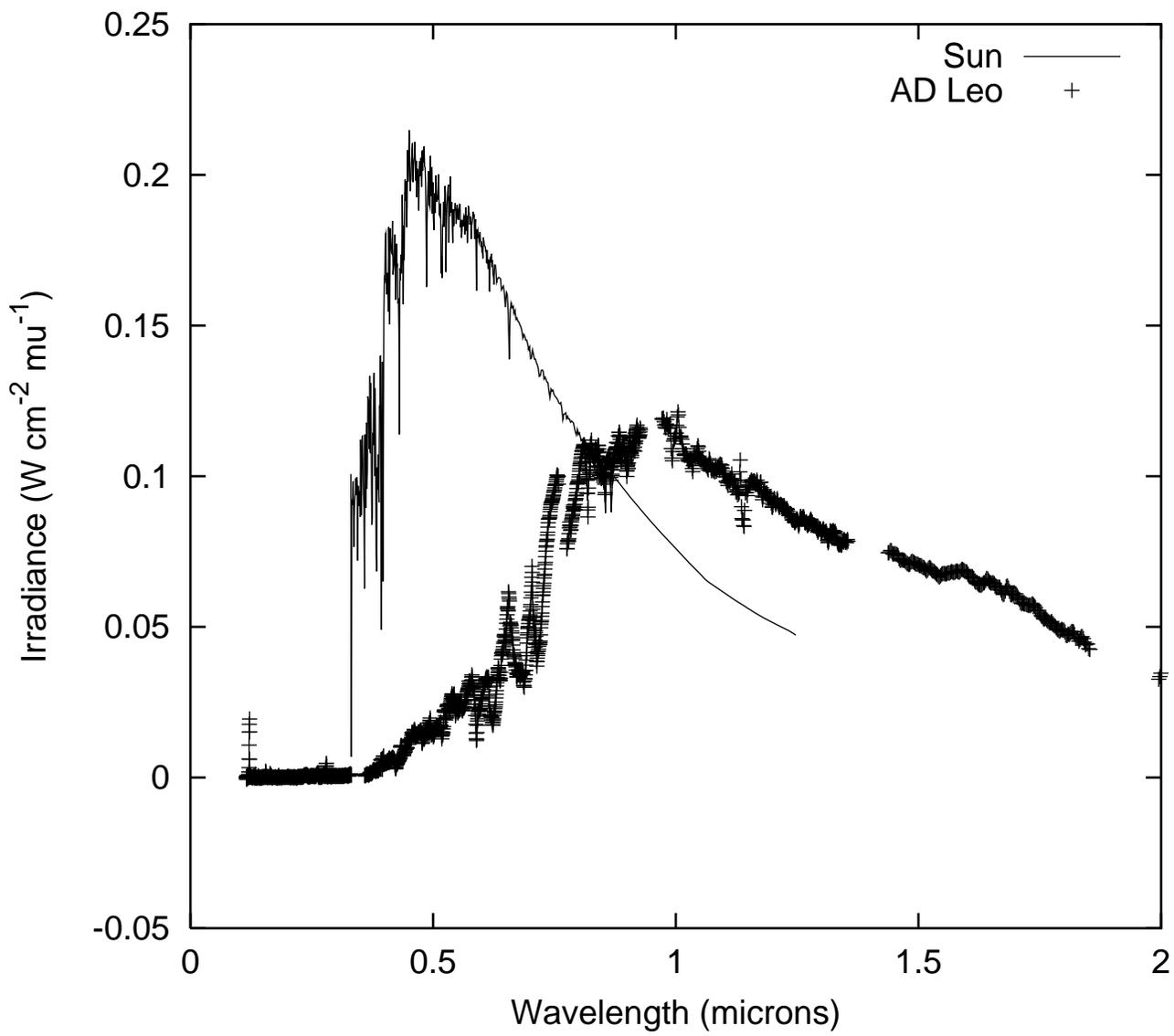

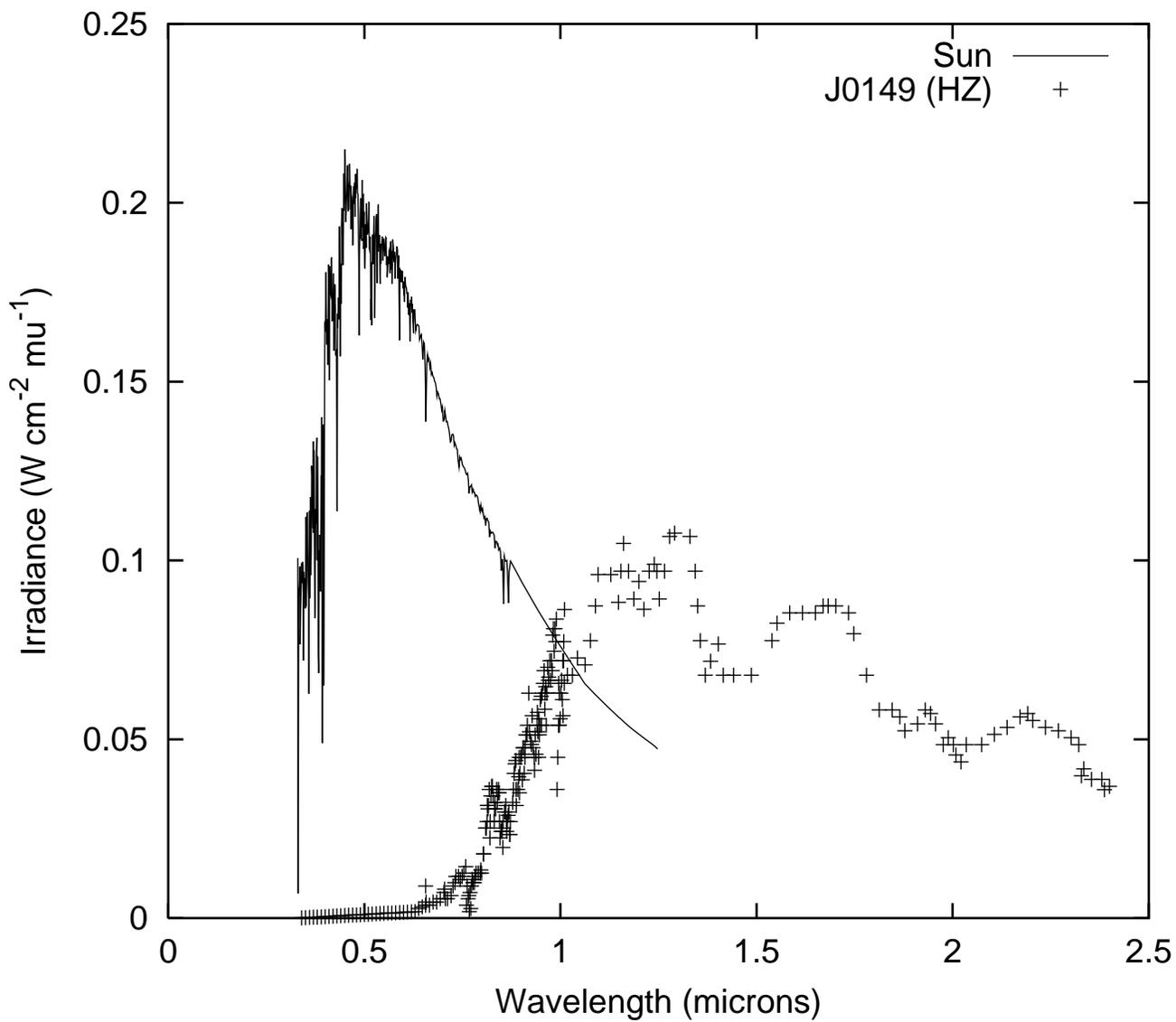

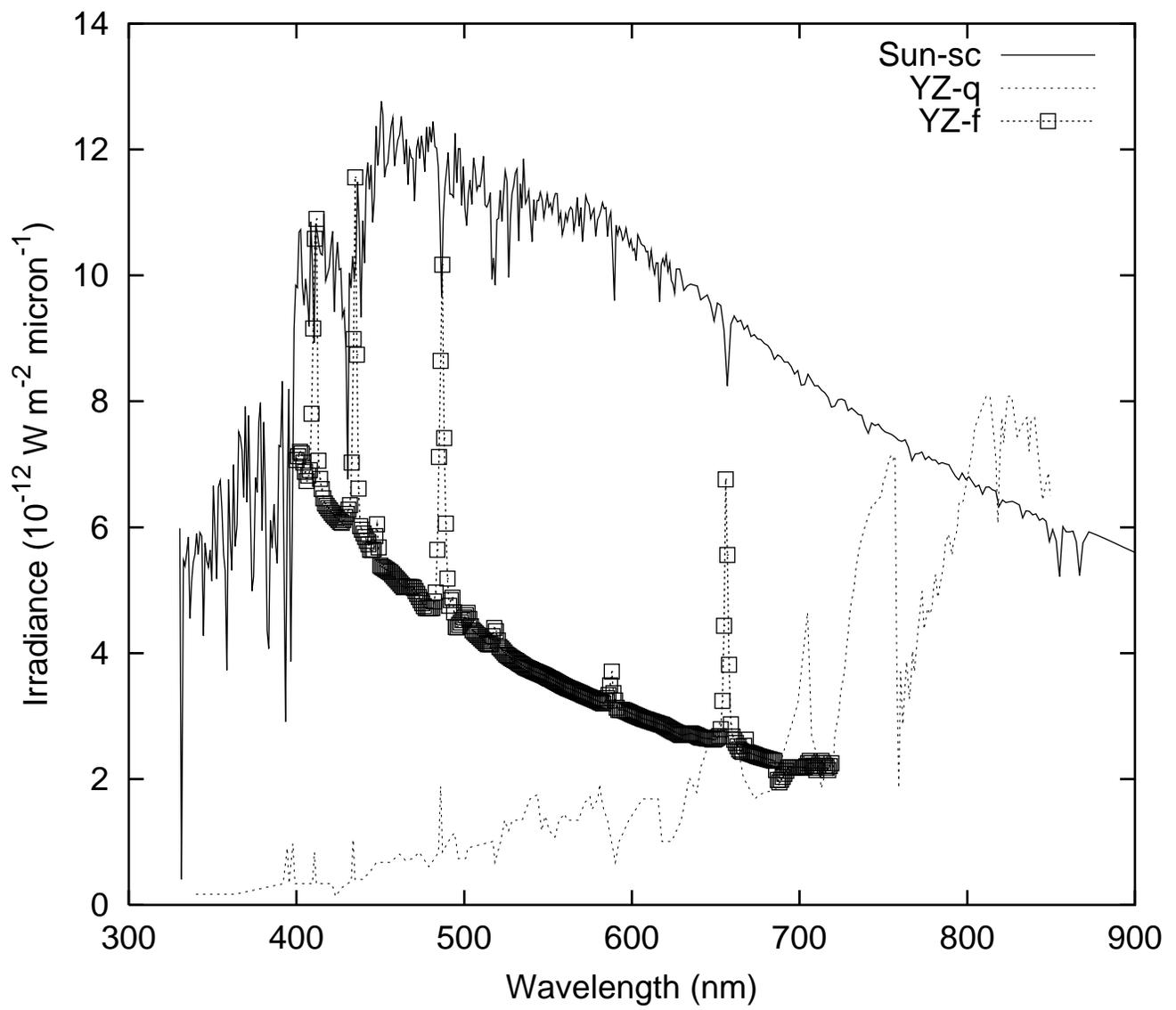

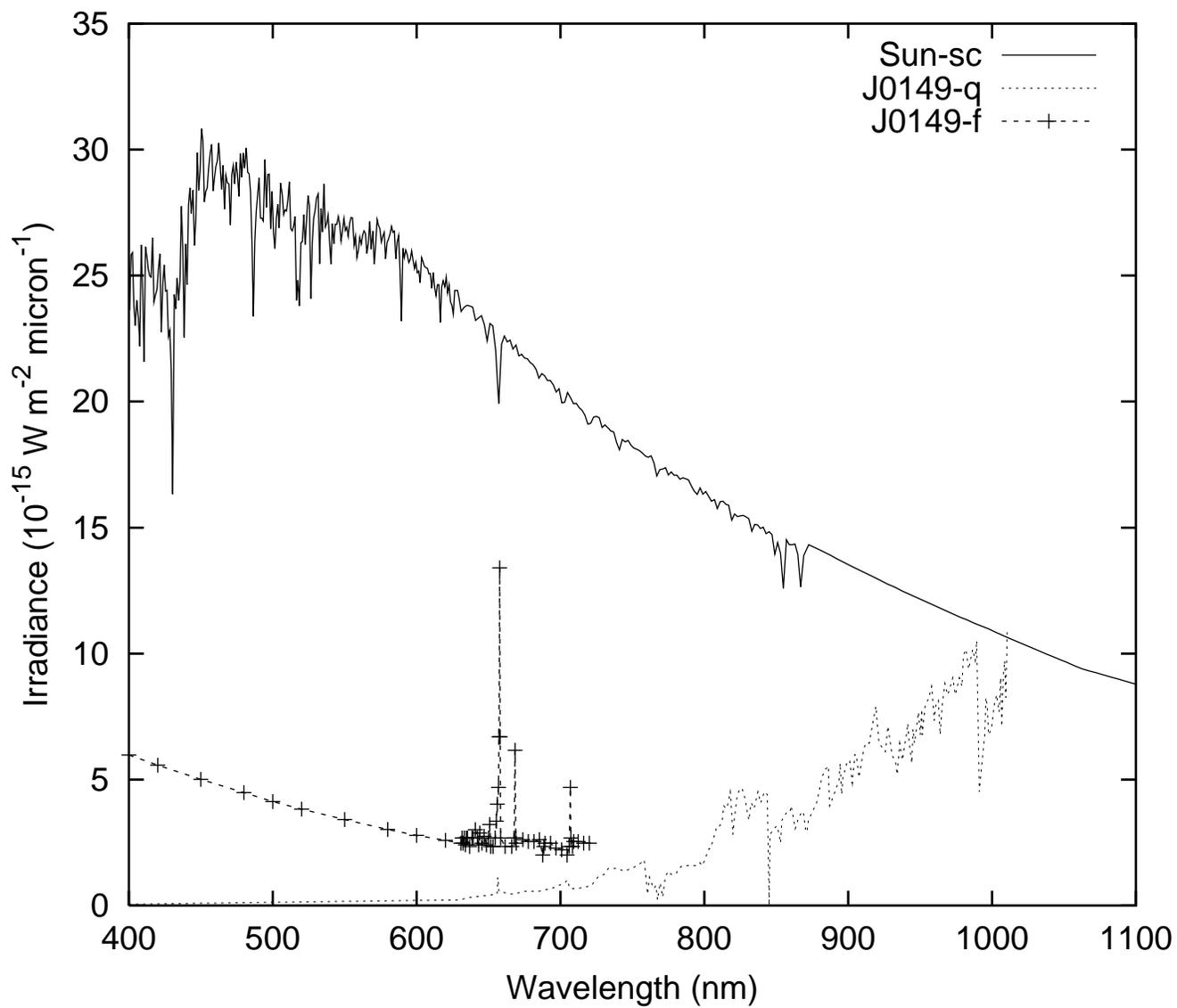

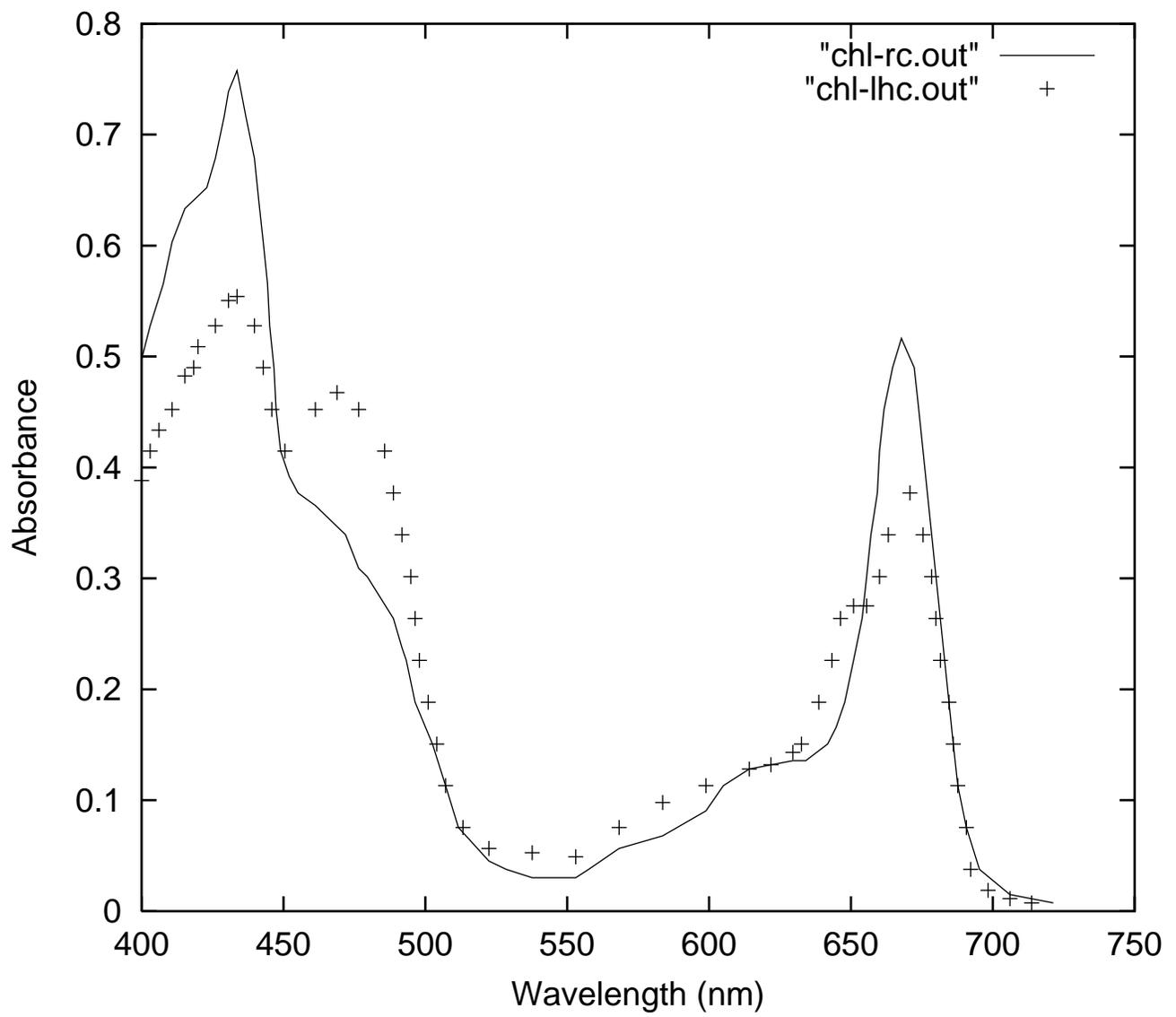